\documentclass[a4paper]{jpconf}
\usepackage{graphicx}
\usepackage[utf8x]{inputenc}
\usepackage{wrapfig}
\usepackage{float}

\begin{document}
\vspace*{-1in}
\title{High Temperature Raman analysis of Hydrogen Phase IV from Molecular Dynamics}
\author{Ioan B Magdău and Graeme J Ackland}
\address{CSEC, SUPA, School of Physics and Astronomy, The University of Edinburgh, Edinburgh EH9 3JZ, United Kingdom}
\ead{i.b.magdau@sms.ed.ac.uk, gjackland@ed.ac.uk}

\begin{abstract}

We extend our previous studies on phase IV of solid hydrogen
by employing larger cells and $k-$sampling. We show that uncorrelated
hexagonal rotations in the weakly bounded G"-layers are needed to account
for the experimentally measured Raman spectrum. In large simulations
we find no evidence of proton diffusion or layer fluctuations, which
we believe are the result of finite size effects. In calculations at
higher pressures (above 400 GPa) we identify a new possible candidate
for phase IV. We, finally, proposed a revised phase diagram based on
our previous and present results.

\end{abstract}

\section{Introduction}

The recent discovery of a high-pressure, high-temperature phase IV of
solid hydrogen \cite{howie2012mixed} has reignited the research
interest in the field. Considerable work has been done to identify the
true nature of this new thermodynamically stable phase. \textit{Ab
  initio} Random Structural Searching (AIRSS) performed at 0 K
\cite{pickard2007structure,pickard2012density} had previously
suggested a new class of mixed crystals, consisting of two types of
layers: molecular (named ``bromine-like'', B) and atomic (named
``graphene-like'', G).  These explained the observation of two
Raman-active vibrons with very different frequency\cite{howie2012mixed, eremets2011conductive}.
Metadynamics simulations reported a similar structure
similar to $Pc$, but with rotational disorder in the B-layer, and
molecular dynamics\cite{magduau2013identification} showed the
possibility of a number of layered structure, stabilised at high
temperature by the entropy of molecular rotations.

Further Molecular Dynamics calculations
attempting to include temperature effects reported unusual behaviour
like "pronounced proton diffusion" \cite{liu2013proton}, "intralayer
structural fluctuations" and "proton tunneling phenomena"
\cite{goncharov2013bonding}. In our recently published work
\cite{magduau2013identification} we showed that these predictions
are an unwanted effect of the finite size simulations and we
offer an alternative interpretation of phase IV.

The mixed cells explain the existence of two Raman vibrons,
\cite{howie2012mixed, eremets2011conductive}. 
and the calculated hard vibron frequency,
arising from the B-layer of free molecules in Pc-structure, is in good
agreement with experiment.  However, the soft vibron, from the interacting
molecules in the G-layer does not agree well when calculated with
Density Functional Perturbation Theory (DFPT) at 0 K. Additionally,
more recent IR measurements at high pressures and room temperature
have found disagreements with the theoretical predictions for phase IV
\cite{eremets2013infrared, loubeyre2013hydrogen, zha2013high}.

We developed a method to extract Raman from Molecular Dynamics at
finite temperature and found that rotation the trimer rings in the
G-layer could account for the hardening of the softer vibron mode
\cite{magduau2013identification}. Here, we validate our Raman method
and also extend our calculations to both larger Pc-cells and larger
$k-$point sampling to gauge the contribution of finite size effects to
our previous simulations.  We also examine the effects of initial
conditions with MD simulations starting from two structures: $Pc$ and $Cc$.

\section{Calculation Details}

\subsection{Validating the molecular projections assumption.}

\begin{wrapfigure}{R}{90mm}
\includegraphics[width=85mm]{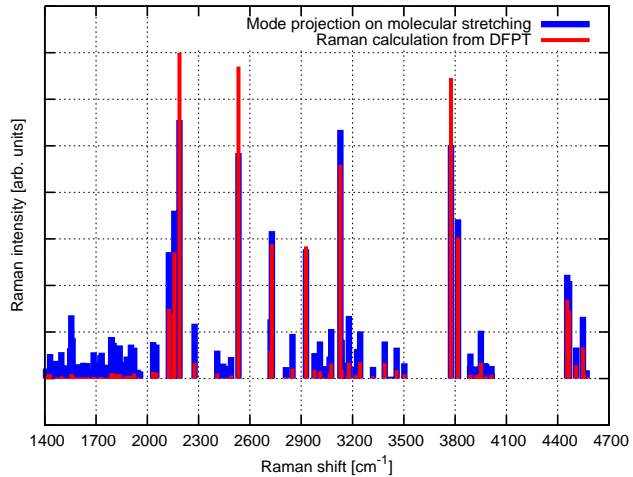}
\caption{Red bars show the Raman intensities calculated with DFPT, blue
  bars show our approach to calculating Raman applied to the
  hydrogen-deuterium crystal by projecting the mode displacements onto
  molecular axis\cite{magduau2013identification}.}
\label{compar}
\end{wrapfigure}

When calculating Raman from MD, our assumption was that Raman activity
can be accounted for by pure molecular stretching from both B and
G-layers. For validating this hypothesis, we investigate a
random hydrogen-deuterium mixture cell (96 atoms, 8x8x8 $k-$points).  We
chose a mixture because it gives us a rich Raman spectrum to compare
against.

We then performed an expensive Raman calculation at the $\Gamma$-point
with DFPT as implemented in CASTEP\cite{segall2002first}, and the
dense $k-$point mesh necessary for convergence of polarisability
calculations.  We also calculated Raman with our alternative method by
projecting all phonon modes onto molecular stretching:
\[ I(\omega_i) = \sum_j \bf{\epsilon_{ij}} \cdot \left(\bf{r_{n_j}} - \bf{r_j}\right) \]
where $\bf{\epsilon_{ij}}$ is the displacement of atom $j$ within mode
$i$ as calculated with Lattice Dynamics (LD), $\bf{r_j}$ is the
position of atom $j$ and $\bf{r_{n_j}}$ is the position of the other
atom in the molecule.

Figure \ref{compar} shows that the two calculations are in excellent agreement, especially for high frequencies (above 2000 $cm^{-1}$).
Although we cannot directly trust the intensities, the positions of the peaks are consistent. This validates
our assumption and allows us to extract Raman from MD by projecting the atomic velocities onto
molecular stretches instead of DFPT phonon modes (the latter becomes unreliable because normal modes
change due to free rotations and molecular flips).

\subsection{Extended MD calculations.}

We have performed systematic MD simulations across a large
pressure-temperature range. Here, we report studies employing larger cells and denser $k-$point sampling than previously.
All DFT calculations were performed using plane waves (PW) with 1200 eV cut-off, ultrasoft potentials
and PBE functional. We start by relaxing our $Pc$ and $Cc$ cells and then carry out MD simulations as
summarised in Table \ref{MD}. We have, additionally, performed MD simulations at 340, 410 and
480 GPa with $Pc-768$ cells, but these exhibit interesting phase transitions,
which will be detailed in the following.
We extract Raman from trajectories using molecular projections\cite{magduau2013identification}:
\[ I(\omega) = FFT \sum_j\bf{v_j}(t)\cdot[\bf{r_{n_j}}(t) - \bf{r_j}(t)]\]
where $\bf{v_j}(t)$ is the velocity of atom $j$ at time $t$ measured starting after an equilibration period, and $FFT$ stands for Fast Fourier Transform.
We also evaluate the atomic square root displacements as:
\[ D(t) = \frac{1}{N} \sum_j \sqrt{\bf{r_j}(t)^2-\bf{r_j}(0)^2} \]

\begin{table}[H]
\caption{\label{MD}Details of MD simulations.}
\begin{center}
\begin{tabular}{lllllll}
\br
Cell&No. Atoms&Layers&Pressure&Temperature&$k-$points&Length/Ensemble\\
\mr
Pc&48&4&250 GPa&325 K&5x3x3&3 ps NVE\\
Pc&288&4&250 GPa&325 K&1x1x2&1.5 ps NVE\\
Pc&288&4&270 GPa&300 K&2x2x4&1.5 ps NPT + 1.5 ps NVE\\
Cc&384&4&270 GPa&300 K&1x1x2&0.3 ps NPT + 2.0 ps NVE\\
Pc&768&8&270 GPa&300 K&1x1x1&0.2 ps NPT + 1.5 ps NVE\\
\br
\end{tabular}
\end{center}
\end{table}

\section{Results and discussion}

\begin{figure}[h]
\centering
\includegraphics[width=150mm]{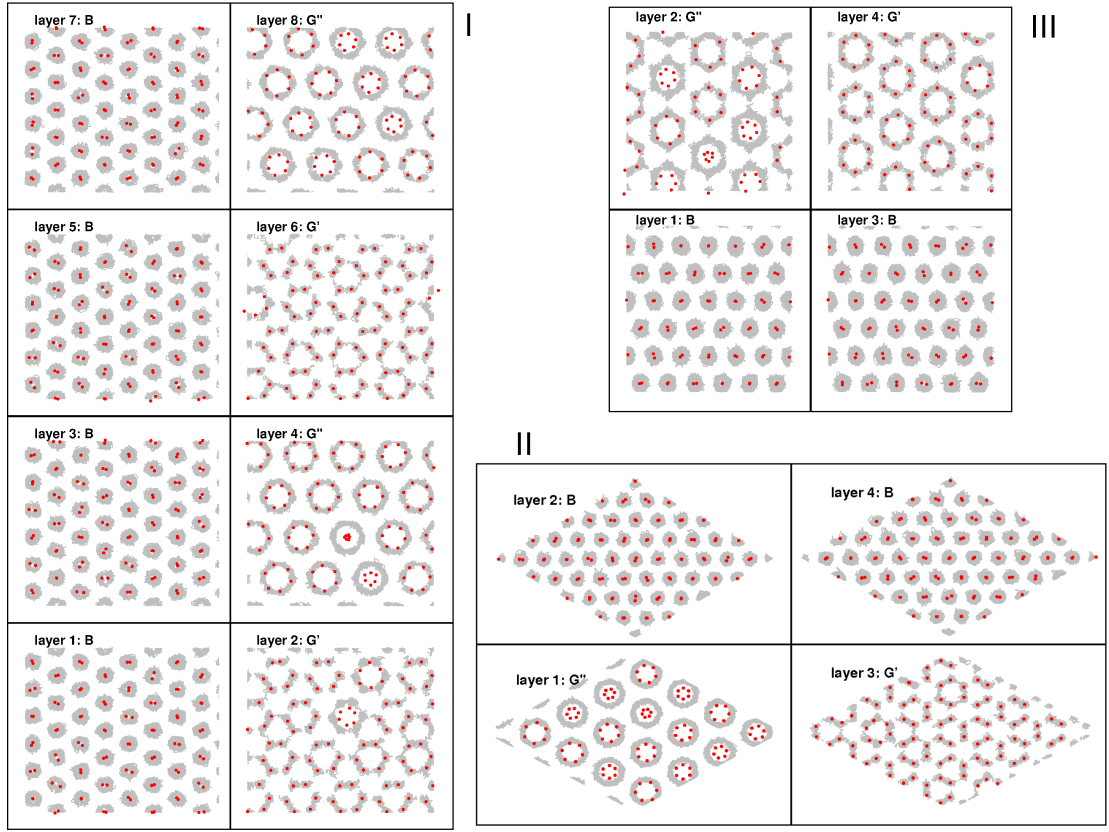}
\caption{The figures shows trajectories (grey) and average atomic
  positions (red dots) extracted from MD simulations at 270GPa and
  300 K. Panel I: Pc, 768 atoms, 1x1x1 $k-$points; Panel II: Cc, 384
  atoms, 1x1x2 $k-$points; Panel III: Pc, 288 atoms, 2x2x4 $k-$points.
  When averaged over a long enough period to allow
  full rotations, the two red dots in the B-layer molecule / six
  red dots in the G'' trimer become coincident at the centre of molecule / thorus.}
\label{cells}
\end{figure}

All extended calculations are in good agreement with our previous results.
As illustrated in Figure \ref{cells}, in the stability regime of phase IV,
the B-layers are freely rotating, while the G-layers differentiate in two distinct
types. The G' layers are stable and while snapshots show distinct molecules with electrons located in covalent bonds,
 on time-averaging the G' layers exhibit hexagonal symmetry.
The strongly-bonded molecules seem to be part of two trimers at the same time,
and molecular flips (out of the plane) are possible, but infrequent. By contrast,
in the G" layers, the trimers are uncorrelated and rotating giving
strong inter-molecular interactions and quick rebonding. The bonds in this
latter layer are characterised by an accentuated anharmonicity.

Our observations are in particular validated by the large 768 atoms simulation 
(Panel I in Figure \ref{cells}). We carefully prepare 8
extended layers, containing 16 trimers each, in the $Pc$ structure. 
Initially, all 4 G-layers have the same
symmetry, but during the course of the simulation, they differentiate
in an alternating stacking of G' and G" layers. We also find that simulations started in the
$Cc$ cell (Panel II, Figure \ref{cells}) proposed by Liu \textit{et al}
\cite{liu2012room} displays a very similar high temperature behaviour. The high temperature structure observed 
seems to be independent of
the choice of initial conditions or unit cell (orthogonal/monoclinic), although
finite size effects may play a role: in our smaller cell with a denser $k-$point mesh
(Panel III, Figure \ref{cells}), the rotation of the hexagons took
longer to initiate and was not consistent throughout the simulation.

Also, our calculation is based on a classical description. Across a range of
systems, the effect of including zero-point energy or using nuclear
wavefunctions is similar to increasing temperature.  Including the
quantum effects of the protons (especially the in-molecule
indistinguishability implied by para-hydrogen) would facilitate the
hexagon rotations in the G" layers, increasing the vibron
frequency and reducing the small disagreements with experiment still
further.

\begin{wrapfigure}{L}{75mm}
\includegraphics[width=70mm]{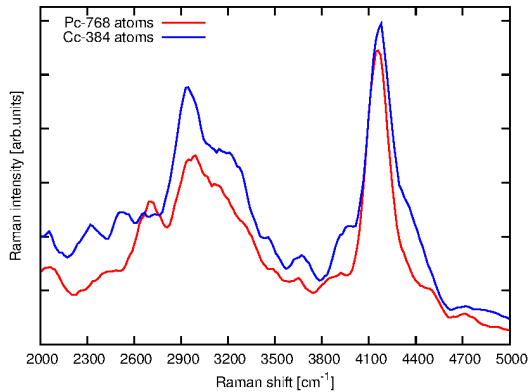}
\caption{We show Raman computed with the molecular projection method for: Pc (768 atoms) (red)
and Cc (384 atoms) (blue). The results are consistent.}
\label{raman}
\end{wrapfigure}

We further compare the Raman spectrum of $Pc-768$ and $Cc-384$ cells 
at 270 GPa and room temperature as shown in Figure \ref{raman}.
In both cases, we obtain two well defined peaks at around 4150
$cm^{-1}$ and 3000 $cm^{-1}$. The results are consistent and in good agreement
with our previous study on smaller systems\cite{magduau2013identification} and also in accordance with the experimental measurements\cite{howie2012mixed}
at the same conditions of pressure and temperature. We find no dependence
on whether the simulation was initiated in $Pc$ or $Cc$.

In the smaller $Pc$ cell with denser $k-$meshing, the softer vibron is
too soft, similar to LD calculation.  In other words, when hexagons are not rotating, the MD
simulation is the same as a simple oscillation around the equilibrium
position, similar with LD. Similarly, we previously reported that at lower temperatures the 
rotation stops and the Raman frequencies soften.\cite{magduau2013identification}
This is further evidence that the rotating-trimer G"
layers describe the true nature of phase IV.

We thus conclude that uncorrelated trimer
rotations in the G" layers are necessary to explain
the measured Raman spectrum.  We expect that quantum effects on the protons
will only enhance the rotation.

The same effect could be responsible for the
hardening of the softer IR peak which is currently the topic of much
debate in the literature. We intend to calculate IR with similar
methods in future studies and confirm that mixed structures are
the best candidates for phase IV of solid hydrogen.

\begin{wrapfigure}{r}{95mm}
\includegraphics[width=90mm]{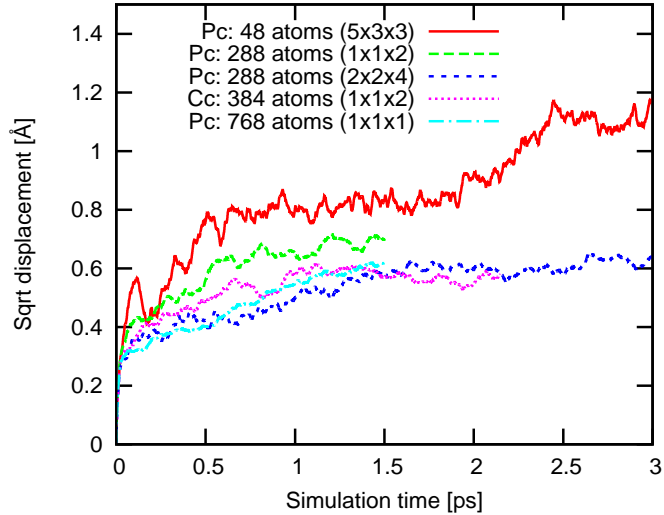}
\caption{Figure shows the square root displacement extracted from
MD calculations at various conditions as summarised in Table
\ref{MD}. Numbers in brackets are the $k$-point set.}
\label{diff}
\end{wrapfigure}

Apart from the Raman spectra, we also use the MD trajectories to
investigate previously-claimed proton 
diffusion \cite{liu2013proton}. We extract the mean
square root displacements (MSD) from the MD
trajectories and plot it against time as shown in Figure \ref{diff}.

It is clear that increasing the size of the cells drastically
reduces the MSD and apparent diffusion.

In the 48-atom cell, various finite size effects, like layer
fluctuations and simultaneous hexagon rotations, combined with unclear
trimer identity, can facilitate effects like proton diffusion.  This is
illustrated by the sudden jumps in the red curve (see Figure
\ref{diff}). However, when larger cells are involved, there is a rapid
increase in MSD which can be explained by the free rotation of the
molecules in the B-layers and uncorrelated rotations of the hexagons
in the G"-layers. The MSD then stabilises at a constant value representing
the radius of the trimer/molecule (grey torus in Fig 2).  This effect is
already present in cells of 288 atoms, which we note is larger than
any simulation for which data is presented in previous
work \cite{liu2013proton, goncharov2013bonding}.

\begin{wrapfigure}{l}{50mm}
\includegraphics[width=45mm]{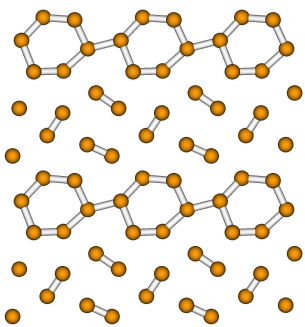}
\caption{A layer of the new ABAB stacked structure discovered at 480 GPa.
The novel cell could possibly be a candidate for phase IV or
another higher-pressure phase.}
\label{new}
\end{wrapfigure}

The rapid proton diffusion which we observe in smaller simulations can
be traced to reconstructions of {\it entire layers}.  The stochastic
probability of such reconstructions drops exponentially with the
number of molecules in each layer, so the proton-diffusion
``mechanism'' is absent in larger simulations, and will be absent in
still-larger experimental samples.  The absence of proton diffusion
can also be observed in the trajectories plotted in Figure
\ref{cells}, in which all protons stay in their
initial B-molecules or G-trimers.

We have already shown that the proton-diffusion mechanism is not
needed to explain the measured Raman spectra as was previously
claimed: the rotation of the trimers is sufficient.

In simulations at higher pressures we find a series of phase transitions.
The slowly-converging relaxation of the $Pc-768$ atoms cell at 480 GPa
finally transformed to a new, unreported structure as shown in
Figure \ref{new}. Although unstable at these conditions,
the new layered cell is interesting because it contains
two different molecular environments within 
the same layer. As seen in other mixed-structures, this is the kind of
setup that could give two Raman vibrational peaks of different
natures. Further shaking and relaxations of the new cell at 0 K resulted
in a transformation to the previously predicted $Cmca-12$ \cite{pickard2007structure}.

\begin{figure}[h]
\begin{minipage}{80mm}
\fbox{\includegraphics[width=80mm]{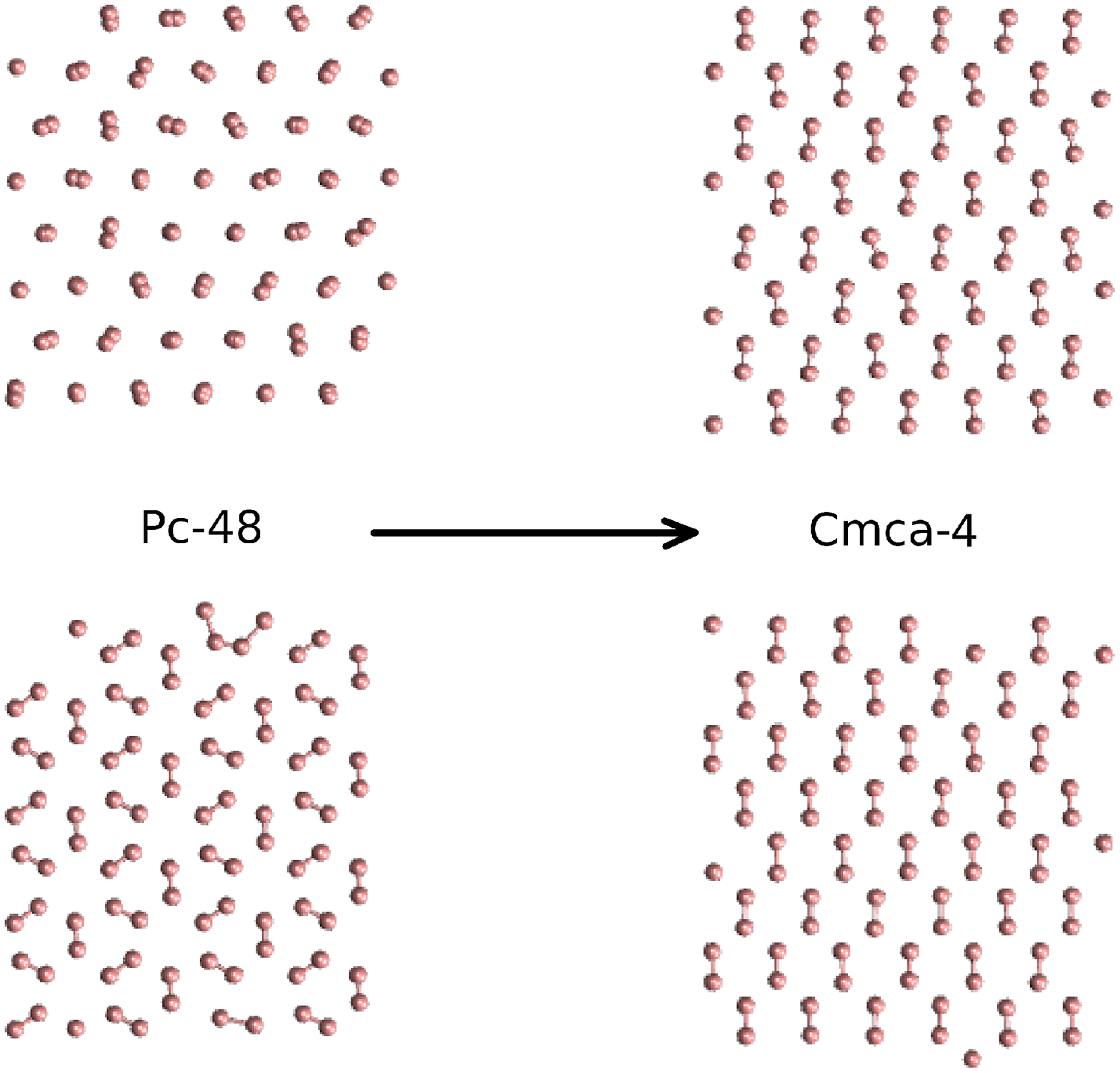}}
\caption{\label{phase}Figure shows average atomic positions extracted from the MD runs
at 340 GPa. The Pc structure (left) transforms to Cmca-4 (right) in less than 0.2 ps of NPT
simulation.}
\end{minipage}
\hspace{2mm}
\begin{minipage}{75mm}
\includegraphics[width=75mm]{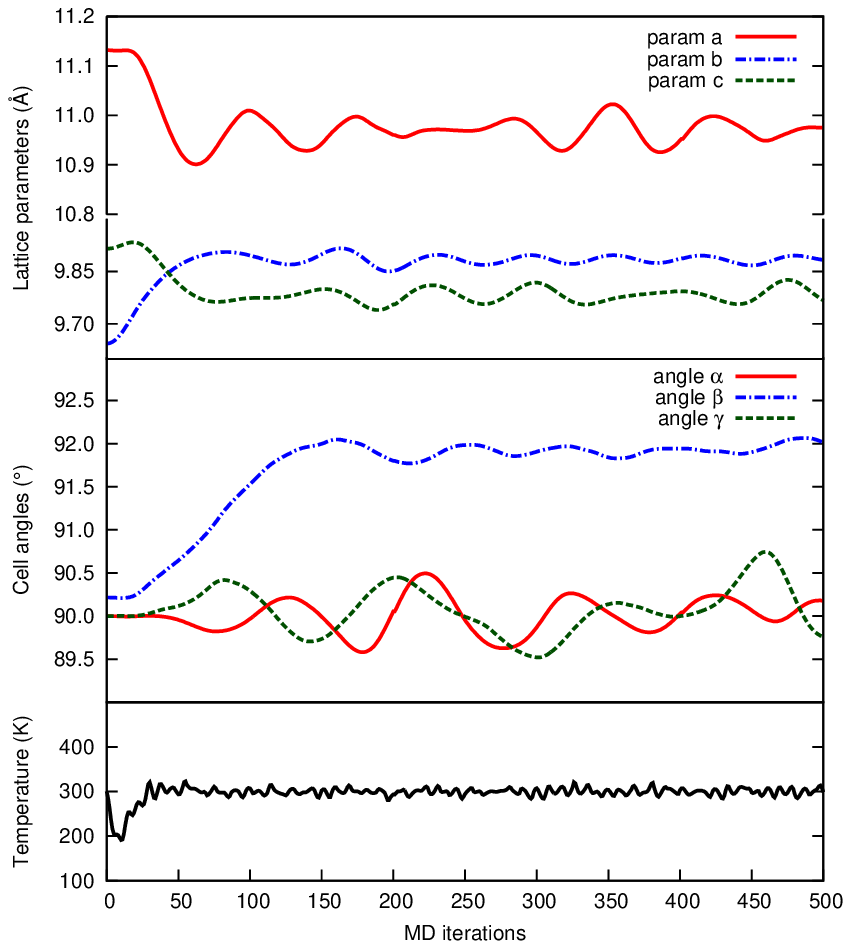}
\caption{\label{param}The Pc to Cmca-4 transition leads to an abrupt
change in the lattice parameters $a$ and $\beta$ as shown in the middle
and top panels.}
\end{minipage}
\end{figure}

\begin{wrapfigure}{r}{90mm}
\includegraphics[width=90mm]{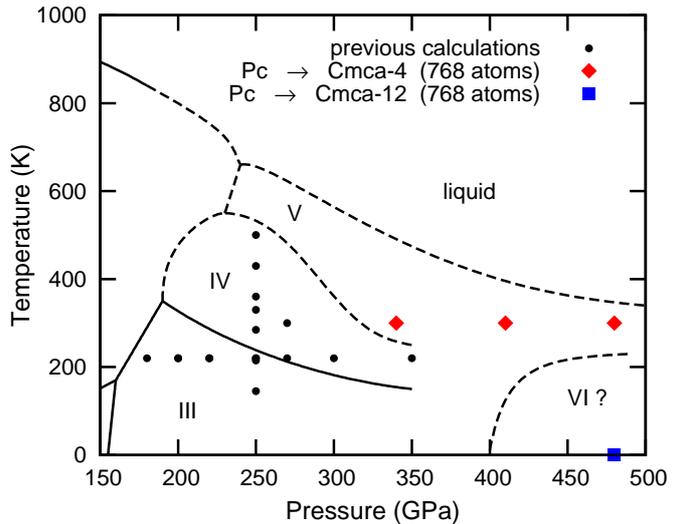}
\caption{Revised Phase Diagram.}
\label{diagram}
\end{wrapfigure}

We started MD simulations with the large cells (768 atoms) using the NPT
ensemble at room temperatures (see Figure \ref{diagram}). After less than
0.2 ps all structures above 340 GPa went through a phase transition as 
shown in Figures \ref{phase} and \ref{param}. The average atomic positions is this new phase are
very similar to Pickard's $Cmca-4$ structure \cite{pickard2012density}. The transition involves
an abrupt change in the lattice parameters, especially cell length $a$
and cell angle $\beta$ as shown in Figure \ref{param}. Since this
phase has only one type of molecular layer, transitions from $Pc$ could
easily be identified in experiments by disappearance of one of the Raman vibrons.

Recent experiments reported by Howie at this meeting,
suggest such Raman transformations induced by heating.  This could be
accounted for by either a liquid phase, or possibly the $Cmca-4$ structure which we have also found
in MD simulations in agreement with previous metadynamics studies \cite{liu2012room}.

Summing up the results from our previous studies and our current work we speculate
a revised phase diagram. We expect that future experiments at both low and room
temperatures will find new phases upon increasing the pressure. We, further, found
no evidence of melting in our MD simulations in agreement with other recent
studies which predict a minimum in the melting curve \cite{liu2013anomalous}.
The possibility of a ground liquid state \cite{bonev2004quantum}
at these pressure is questionable.

\section{Conclusions}

In conclusion, we have validated our previous studies
\cite{magduau2013identification} by extending the simulations to
larger cells and denser $k-$meshes. We have also checked our simulations 
against other groups'\cite{liu2013proton, goncharov2013bonding}. 

Specifically, we have shown that the layer reconstruction, which 
can lead to linearly-increasing MSD and ``proton diffusion'', is a finite size
effect. 
We found no evidence of such a phenomenon in larger
cells on the simulated time scale. Uncorrelated rotations of trimers
explain the Raman spectra as observed in experiments and are most
probably the cause of similar changes in IR vibrons. 
Possible candidates for phase V of hydrogen (Cmca-4) and VI (Cmca-12) might become
stable at certain conditions. Finally, we speculate a new, revised
phase diagram based on our old and present calculations.

\ack

We acknowledge E. Gregoryanz for numerous useful
discussions and EPSRC for a studentship (I.B.M.).

\section*{References}

\bibliographystyle{iopart-num}
\bibliography{References}

\providecommand{\newblock}{}
\begin{thebibliography}{10}
\expandafter\ifx\csname url\endcsname\relax
  \def\url#1{{\tt #1}}\fi
\expandafter\ifx\csname urlprefix\endcsname\relax\def\urlprefix{URL }\fi
\providecommand{\eprint}[2][]{\url{#2}}

\bibitem{howie2012mixed}
Howie R~T, Guillaume C~L, Scheler T, Goncharov A~F and Gregoryanz E 2012 {\em
  Physical Review Letters\/} {\bf 108} 125501

\bibitem{pickard2007structure}
Pickard C~J and Needs R~J 2007 {\em Nature Physics\/} {\bf 3} 473--476

\bibitem{pickard2012density}
Pickard C~J, Martinez-Canales M and Needs R~J 2012 {\em Physical Review B\/}
  {\bf 85} 214114

\bibitem{eremets2011conductive}
Eremets M and Troyan I 2011 {\em Nature materials\/} {\bf 10} 927--931

\bibitem{magduau2013identification}
Magd{\u{a}}u I~B and Ackland G~J 2013 {\em Physical Review B\/} {\bf 87} 174110

\bibitem{liu2013proton}
Liu H and Ma Y 2013 {\em Physical review letters\/} {\bf 110} 025903

\bibitem{goncharov2013bonding}
Goncharov A~F, John S~T, Wang H, Yang J, Struzhkin V~V, Howie R~T and
  Gregoryanz E 2013 {\em Physical Review B\/} {\bf 87} 024101

\bibitem{eremets2013infrared}
Eremets M, Troyan I, Lerch P and Drozdov A 2013 {\em High Pressure Research\/}
  1--4

\bibitem{loubeyre2013hydrogen}
Loubeyre P, Occelli F and Dumas P 2013 {\em Physical Review B\/} {\bf 87}
  134101

\bibitem{zha2013high}
Zha C~s, Liu Z, Ahart M, Boehler R and Hemley R~J 2013 {\em Physical Review
  Letters\/} {\bf 110} 217402

\bibitem{segall2002first}
Segall M, Lindan P~J, Probert M, Pickard C, Hasnip P, Clark S and Payne M 2002
  {\em Journal of Physics: Condensed Matter\/} {\bf 14} 2717

\bibitem{liu2012room}
Liu H, Zhu L, Cui W and Ma Y 2012 {\em The Journal of Chemical Physics\/} {\bf
  137} 074501

\bibitem{liu2013anomalous}
Liu H, Hernandez E~R, Yan J and Ma Y 2013 {\em The Journal of Physical
  Chemistry C\/}

\bibitem{bonev2004quantum}
Bonev S~A, Schwegler E, Ogitsu T and Galli G 2004 {\em Nature\/} {\bf 431}
  669--672

\end{thebibliography}

\end{document}